\documentclass[conference]{IEEEtran}
\IEEEoverridecommandlockouts
\usepackage{amsmath,amssymb}
\usepackage{graphicx}
\usepackage{braket}
\usepackage{epsfig}
\usepackage{epstopdf}
\usepackage{mdwmath}
\usepackage{xcolor}
\usepackage{varwidth}
\usepackage{framed}
\usepackage{float}
\usepackage{placeins}
\usepackage{afterpage}
\usepackage{ragged2e}
\usepackage{amssymb}
\usepackage{pifont}

\usepackage{blindtext}
\usepackage{enumitem}
\usepackage{xcolor}
\usepackage{eqparbox}
\usepackage{fixltx2e}
\usepackage{multirow}
\usepackage{enumitem,lipsum}

\usepackage[english]{babel}
\newcolumntype{P}[1]{>{\centering\arraybackslash}p{#1}}
\usepackage{graphicx}
\usepackage{amsmath, amssymb}
\usepackage{graphicx}
\usepackage{epsfig}
\usepackage{epstopdf}
\usepackage{array}
\usepackage{graphicx}
\usepackage{subcaption}
\usepackage{hyphenat}

\usepackage{booktabs} 
\usepackage{graphicx}
\usepackage{caption}
\usepackage{bbding}
\usepackage{pifont}
\usepackage{wasysym}
\usepackage{amssymb}

\usepackage{algorithm}
\usepackage{algorithmic}

\usepackage{framed,url,caption,graphicx}
\usepackage{soul,xspace,tabularx}

\expandafter\def\expandafter\UrlBreaks\expandafter{\UrlBreaks
  \do\a\do\b\do\c\do\d\do\e\do\f\do\g\do\h\do\i\do\j%
  \do\k\do\l\do\m\do\n\do\o\do\p\do\q\do\r\do\s\do\t%
  \do\u\do\v\do\w\do\x\do\y\do\z\do\A\do\B\do\C\do\D%
  \do\E\do\F\do\G\do\H\do\I\do\J\do\K\do\L\do\M\do\N%
  \do\O\do\P\do\Q\do\R\do\S\do\T\do\U\do\V\do\W\do\X%
  \do\Y\do\Z}

\providecommand{\myparab}[1]{\smallskip\noindent\textbf{#1} }

\newcommand{\eat}[1]{}

\usepackage{braket}
\usepackage[utf8]{inputenc}


\usepackage[utf8]{inputenc}
\usepackage[english]{babel}

\usepackage{lastpage}

\usepackage{indentfirst}
\usepackage[T1]{fontenc}
\usepackage[utf8]{inputenc}
\usepackage{authblk}
\usepackage{graphicx}

\begin{document}

\title{Robust Path Selection in Software-defined WANs using Deep Reinforcement Learning}

\author{Shahrooz Pouryousef\thanks{shahrooz@cs.umass.edu}}
\author{Lixin Gao\thanks{}}
\author{Don Towsley\thanks{}}
\affil{University of Massachusetts Amherst}

\renewcommand\Authands{ and }

\maketitle
\begin{abstract}
In the context of an efficient network traffic engineering process
where the network continuously measures a new traffic matrix and updates the set of paths in the network, an automated process is required to \textit{quickly and efficiently identify when and what set of paths should be used}. Unfortunately, the burden of finding the optimal solution for the network updating process in each given time interval is high since the computation complexity of optimization approaches using linear programming  increases significantly as the size of the network increases. In this paper, we use deep reinforcement learning to derive a data-driven algorithm that does the path selection in the network considering the overhead of route computation and path updates. Our proposed scheme leverages information about past network behavior to identify a set of robust paths to be used for multiple future time intervals to avoid the overhead of updating the forwarding behavior of routers frequently. We compare the results of our approach to other traffic engineering solutions through extensive simulations across real network topologies.  Our results demonstrate that our scheme fares well by a factor of $40\%$ with respect to reducing link utilization compared to traditional TE schemes such as ECMP. Our scheme provides a slightly higher link utilization (around 25\%) compared to schemes that only minimize the link utilization and do not care about path updating overhead.

\end{abstract}

\begin{IEEEkeywords}
Network updates, Traffic Engineering, Path selection, Reinforcement Learning
\end{IEEEkeywords}

\section{Introduction}

Network updates for traffic engineering (TE) is a difficult online decision-making problem. In SDN-based networks, a centralized controller usually solves an optimization problem to first compute the new forwarding paths in order to minimize or maximize an objective function for each time interval of route computation \cite{kumar2018semi,abuzaid2021contracting}. Then, the new forwarding rules are installed on routers through a consistent network update process (i.e., enforces TE policies via add, remove and modify forwarding rules) to avoid loops and congestion in the network \cite{jin2014dynamic,reitblatt2012abstractions,khurshid2013veriflow,kim2015kinetic,horn2017delta}. Each or both of these steps can be formulated as a linear programming (LP) optimization problem to find the optimal solution.

Unfortunately, the burden of finding the optimal solution for the network update process is high \cite{abuzaid2021contracting,zhu2021network,kumar2018semi,jin2014dynamic,reitblatt2012abstractions,khurshid2013veriflow}. The key challenge of the combinatorial optimization problem approaches such as LP is scalability \cite{abuzaid2021contracting,zhu2021network} as  route computation complexity increases with network size. In addition, these approaches require an accurate model of the network state such as network traffic demands or the probability of link failures that usually change over time and are hard to measure. Oblivious approaches \cite{racke2002minimizing,racke2008optimal,hajiaghayi2007semi} avoid the overhead of running an LP in each time interval; but they may suffer a performance penalty as their solution is prone to not match the current/true demand matrix in run-time \cite{bogle2019teavar,benson2010case}.

Recently, there have been attempts to apply machine learning techniques to a wide range of networking problems, specifically routing and TE \cite{valadarsky2017learning,xu2018experience,geng2020multi,zhang2020cfr,sun2020scalable,almasan2022enero}. Most of the existing works only target the route computation process \cite{xu2018experience}, not the whole process of network updates. The process of route computation and path updating process  in the network can be formulated as an online learning problem where the cost of path updates and other objectives such as link utilization are considered jointly \cite{zheng2021online}. However, a common critique of online learning approaches is that they lead to algorithms optimized for worst-case instances. As a result, such algorithms can under-perform for typical instances present
in real-world applications. 

In this paper, we use deep reinforcement learning (DRL) techniques to derive a data-driven algorithm that performs the path selection in the network considering the overhead of route computation and path update process. DRL is a sub-field of machine learning that helps an agent to learn to achieve goals in a complex, uncertain environment. Motivated and inspired by DRL results in online decision-making fields \cite{mnih2013playing,chen2018auto}, our scheme uses DRL to identify when  and how to update the routes in the network considering the current state of the network and the traffic demands. Our proposed scheme for the network update process leverages information about past network behavior to identify a set of robust paths to be used in multiple future time intervals to avoid the overhead of frequent updates of the forwarding behavior of routers  and at the same time guarantees  low link utilizations.

We evaluate and compare our approach to other TE schemes using datasets from real network topologies. We experimentally show that DRL can be used to reduce path update overhead in each TE time interval effectively without increasing link utilization much. Our results confirm that DRL can reduce maximum link utilization by about 40\% compared to classical schemes such as ECMP (Equal Cost Multi-Path). We show that the resulting link utilization in the network in our scheme is around 25\% higher than the greedy approach that solves the multi-commodity flow problem irrespective of the cost of path installation in the network. ECMP and oblivious routing schemes have 55\% higher \textbf{MLU} (\textbf{M}aximum \textbf{L}ink \textbf{U}tilization) compared to the scheme that minimizes the MLU (multi-commodity flow problem solution). 

\section{Background}
\vspace{-0.03in}
In this section, we overview reinforcement learning (RL) and explain how we use deep learning for RL.

\subsection{Reinforcement learning}

\vspace{-0.02in}
Over the past few years, RL has become increasingly popular due to its success in addressing challenging sequential decision-making problems in complex, uncertain environments. In RL, time is divided into time slots $t = 1, 2, 3,...$, and an agent repeatedly interacts with an environment. At the beginning of each time slot $t$, the agent observes the current state $s_{t-1}$ of the environment and selects an action from a set of actions. Once the agent executes the action $a_t$, the state of the environment changes to $s_t$ and the agent receives a real-valued reward, which shows how good the chosen action was. The goal of the agent is to learn a policy $\pi$ that is a mapping from the set of possible states $S$ to the space of actions $A$ that maximizes the expected discounted reward. In general, the objective of RL agent is to maximize the discounted cumulative reward $\sum_{t=1}^T \gamma r_t$ where $\gamma \in [0, 1]$ is the discount factor.

In deep RL, the policy function $\pi$ is approximated with a neural network \cite{arulkumaran2017deep}. In this paper, we use a method called policy gradients \cite{sutton1999policy}, where our simple neural network learns a policy to select actions by adjusting its weights through gradient descent. Policy gradient algorithms typically proceed by sampling
this stochastic policy and adjusting the policy parameters
in the direction of greater cumulative reward \cite{silver2014deterministic}.

\section{Robust path selection using RL}
In this section, we describe the design and implementation of our system. We start by explaining the problem formulation when we know the future traffic matrices and show how to find a set of robust paths to be used for the next $w$ time intervals. Then, we explain the main components and the training algorithm of our DRL-based scheme.

\myparab{Commitment and look-ahead windows:} We use variable $c$ to indicate the commitment window which is the size of the list of traffic matrices in the state of the agent. For each given time $t$, we will use traffic matrices $DM_{t-c},DM_{t-c+1},...,DM_{t-c+c}$ where $DM_t$ is the traffic demand matrix for time interval $t$. $w$ indicates the look-ahead window and indicates the number of next time intervals that the set of suggested paths by the DRL agent would be used for.
\subsection{Optimal robust path selection}
\label{sec:optimal_solution}
In this section, we formulate the problem of optimal path selection for the next $w$ time intervals. Given network topology and DM for the next $w$ time intervals, what set of paths should be used in the network during those time intervals in order to minimize average link utilization?

We use variable $x_p \in \{0,1\}$ to indicate whether path $p$ has been selected to be used in the future $w$ time intervals or not. Variable $r^{(s,d)}_{p,t}$ indicates the rate of traffic on path $p$ for user pair $(s,d)$ at time interval $t$. $D^{(s,d)}_t$ indicates the amount of traffic from source $s$ to destination $d$ at time $t$. The problem formulation is as follows:

\begin{align} 
\label{cons:optimal1}
\min_{r^{(s,d)}_{p,t},x_p}  & \quad \frac{1}{w}\sum_{t\in [t,...,t+w]}Z_{t} 
\\
\text{subject to}\quad
& \quad \forall{t \in [t,...,t+w]:} \nonumber \\
\label{cons:rate_constraint}
\quad & \sum_{p \in P^{(s,d)}}
r^{(s,d)}_{p,t}x_p=1 \quad \forall{(s,d)\in K}\\
\quad & \sum_{\substack{\forall{(s,d) \in K}\\
{p \in P^{(s,d)}}|e \in p}} r^{(s,d)}_{p,t}D^{(s,d)}_tx_p \leq Z_{t} 
\label{cons:link_utilization_constraint}
\quad \forall{e \in E}  \\
\quad &  \sum_{(s,d) \in K\& p \in P^{(s,d)}} x_p \leq |R| \label{cons:set_size_constraint}\\
\quad & r_{p,t}^{(s,d)}\geq0  \quad  \forall{(s,d) \in K}, {p\in P^{(s,d)}}
\label{cons:variable_constraint1}\\
\quad & x_{p}\in \{0,1\}  \quad \forall{(s,d) \in K} \& p \in P^{(s,d)}
\label{cons:variable_constraint2}
\end{align} 

\noindent where $Z_t$ is the expected maximum link utilization for time interval $t$. Constraint (\ref{cons:rate_constraint}) indicates that the sum of the rates on all paths for each user pair $(s,d)$ at each time interval $t$ should be 1. $P^{(s,d)}$ is the set of all paths between user pair $(s,d)$. The left side of constraint (\ref{cons:link_utilization_constraint}) is the link utilization for link $e$ and we want to minimize the upper bound of the link utilization at time $t$ which is $Z_t$. Constraint (\ref{cons:set_size_constraint}) indicates that the sum of all used paths in the next $w$ time intervals should be less than or equal to $|R|$. The value of $|R|$ indicates the maximum number of paths that should be used in the network and set $R$ is the set of robust paths. The size of set $R$ ($|R|$) should be larger than or equal to the number of flows in the network as we want to have at least one path for each flow. The size of set $R$ can be the average number of paths that are used in the network when the routes are computed from the solution of the multi-commodity flow problem. The last two constraints, ((\ref{cons:variable_constraint1}) and (\ref{cons:variable_constraint2})), indicate that the rate of each path at each time interval should be non-negative and that the value of $x_p$ for each path $p$ can be either zero (not selected) or 1 (selected).

Since the optimization problem ($1$) is a non-convex binary optimization, we can not find the optimal solution. For that, we relax the constraints and solve the following optimization. Note that solving this optimization problem and using it in the network is not practical for two reasons. First, we do not know the future demands. Second, it does not scale similarly to solving the multi-commodity flow problem.

In the optimization problem ($1$), if the set of paths is given, then we can efficiently compute the traffic splitting ratios of flows on outgoing links of each node \cite{kumar2018semi}. One way to approximately find the set $R$ is as follows. Since we assume the optimal scheme knows the future traffic demands, we first solve the multi-commodity flow problem for each time interval $t$ in set $[t,...,t+w]$. Then, we use $|R|$ most used paths among them as our robust paths. We use $\tilde{P}^{(s,d)}$ to indicate this set of paths. The below LP formulation (\ref{cons:optimal2}) with the same constraints in (\ref{cons:optimal1}) is used to minimize MLU by
balancing traffic over the allowed base set of paths ($\tilde{P}^{(s,d)}$).:

\begin{align} 
\label{cons:optimal2}
\min_{r^{(s,d)}_{p,t}}  & \quad \frac{1}{w}\sum_{t\in [t,...,t+w]}Z_{t} 
\\
\text{subject to}\quad
& \quad \forall{t \in [t,...,t+w]:} \nonumber \\
\label{cons:rate_constraint2}
\quad & \sum_{p \in \tilde{P}^{(s,d)}}
r^{(s,d)}_{p,t}=1 \quad \forall{(s,d)\in K}\\
\quad & \sum_{\substack{\forall{(s,d) \in K}\\
{p \in \tilde{P}^{(s,d)}}|e \in p}} r^{(s,d)}_{p,t}D^{s,d}_t \leq Z_{t} \quad \forall{e \in E}  \\
\quad & r_{p,t}^{(s,d)}\geq0  \quad  \forall{(s,d) \in K}, {p\in \tilde{P}^{(s,d)}}
\label{cons:variable_constraint12}
\end{align} 


\subsection{Path selection as a DRL problem}
\vspace{-0.03in}
\begin{figure}
\centering
\includegraphics[scale=0.3]{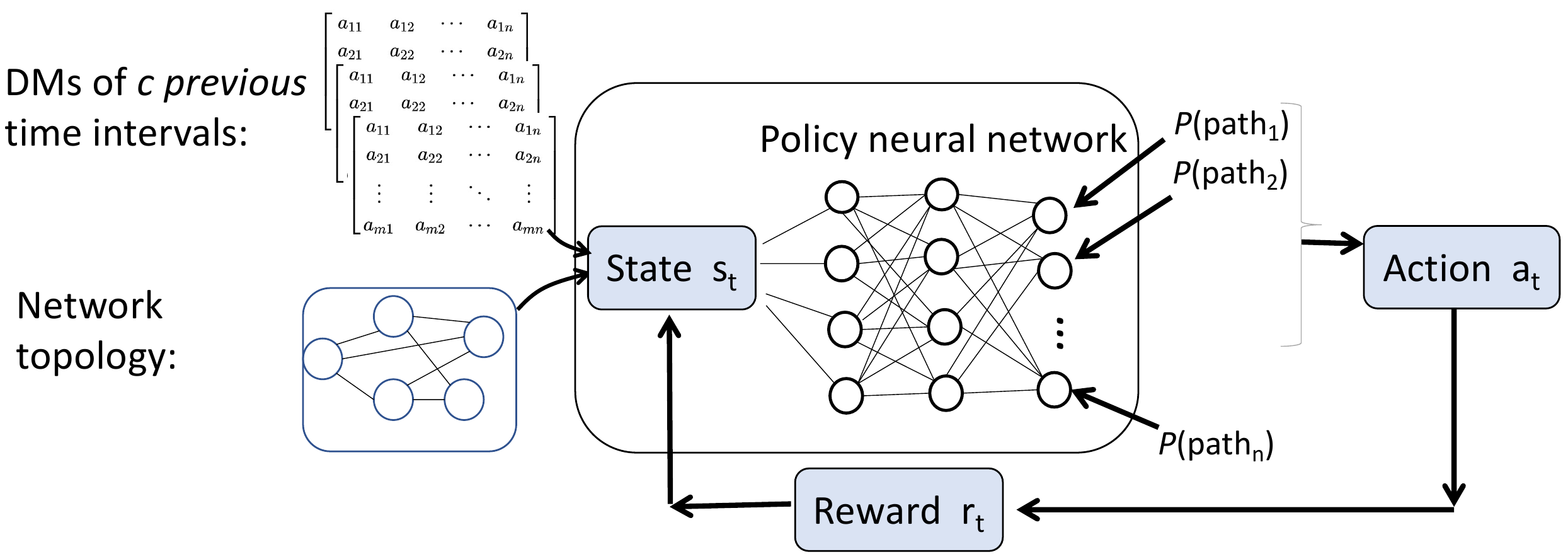}
\vspace{-0.06in}
\caption{Using DRL for path selection in the network.}
\label{fig:architecture}
\end{figure}

Figure \ref{fig:architecture} summarizes how DRL can be applied to path selection in TE. As shown, path selection is derived from training a neural network. The DRL agent observes a set of metrics including the previous $c$ traffic matrices (and possibly other information such as link failure probability) and then it outputs the probability of selecting each path. We use the suggested paths by the DRL agent as input to the problem (\ref{cons:rate_constraint2}) and compute the average maximum link utilization in the network for the next $w$ time intervals. The agent uses the computed reward (explained below) to train and improve its neural network model. We provide more detail about the training algorithm in Section $\S$\ref{sec:training_algorithm}. The state, action, and reward of the DRL agent are as follows:

\subsubsection{State space}In our approach, at each given time $t$, the DRL agent observes the DMs and network topologies for the previous $c$ time intervals.

\subsubsection{Action space}For each state $s_t$, the agent feeds the value of the previous parameters to the neural network and outputs the action, i.e., the probability of selecting different paths. The top $|R|$ paths with the highest probabilities would be used for the next $w$ time intervals.

\subsubsection{Reward}For each set of suggested paths by the DRL agent, we solve the optimization problem (\ref{cons:optimal2}) to find the best rates in each $w$ time interval and measure the average MLU across $w$ time intervals which is $\frac{1}{w}\sum_{t\in [t,...,t+w]}Z_{t} $. We use $\frac{w}{\sum_{t\in [t,...,t+w]}Z_{t} }$ as our reward value.

\myparab{Safe learning:} A solution suggested by DRL agent is unsafe if there is not at least one path for a flow in the network. In order to avoid unsafe solutions, we use safe learning techniques for the online training process proposed in \cite{mao2019towards}. For example, we check if each flow has at least one path in the chosen action by the DRL agent. If not, we add the shortest path for that flow to the set of paths by DRL agent.

\begin{figure}
\centering
\caption{Training Algorithm}
\includegraphics[scale=0.45]{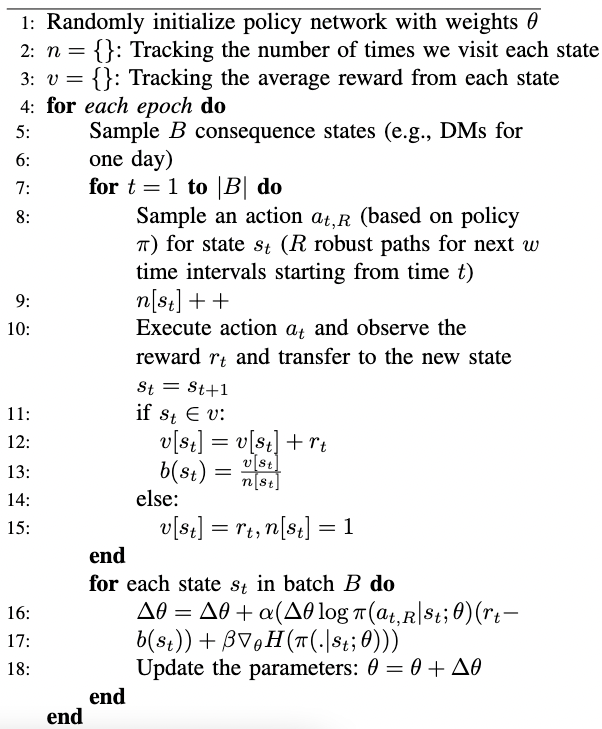}

\label{fig:policy_nework}
\end{figure}
\subsection{Training Algorithm}
\label{sec:training_algorithm}
\vspace{-0.02in}
We now describe our training algorithm. We use a policy gradient method \cite{sutton1999policy} in our algorithm to train the policy of the DRL agent. First, the algorithm randomly initializes all the weights $\theta$ of the policy network; (line 1). The DRL agent updates these parameters over $T$ epochs. For each state $s_t$ that is the matrix of demands for previous $c$ time intervals, it selects action $a_{t,R}$ based on a policy $\pi$ that here is the output of a neural network. Since $|R|$ different paths are sampled for each state $s_t$, we define a solution $a_{t}^{|R|}=(a_{t}^{1} , a_{t}^{2} ,...,a_{t}^{|R|})$ as a combination of $|R|$ sampled actions using a stochastic policy $\pi(a_{t}^{|R|}|s_t)$ parameterized by $\theta$.  We execute the action  and compute the reward and add the state, action, and reward  to the set $B$ to update the weight of the network based on them (lines 8-13). 

For each state $s_t$ in the batch $B$, we use the average reward of the state (computed at line 13) and then use it to update the weights of the neural networks. In lines 16-17, we approximate a stochastic policy $\pi(a_{t} |s_t)$ parameterized by $\theta$ for selecting a solution $a_{t}$ for a given state $s_t$. The approximation is  as follows:

$\pi (a_{t_{|R|}}|s_t;\theta) \approx \prod_{i=1}^{|R|} \pi(a^{i}_{t}|s_t;\theta)$

In order to maximize the expected reward $\mathbb{E}{r_t}$, we use gradient ascend using REINFORCE algorithm \cite{sutton1999policy,silver2014deterministic} with a baseline $b(s_t)$. For each state $s_t$, we use an average reward as the baseline. Other approaches such as ECMP can be considered as our baseline scheme. The policy parameter $\theta$ is updated with learning rate $\alpha$ as following:

$\theta = \theta +\alpha \sum_{t}^{} \nabla_\theta \log \pi(a_{t_{|R|}}|s_t; \theta)(r_t-b(s_t))$.

In the above equation, for a given state $s_t$, we use $(r_t - b(s_t))$ to check how much better a specific solution is compared to the average
solution (our baseline solution). When $r_t > b(s_t)$, the probability of the solution $a_{t_{|R|}}$ is increased by updating policy parameters $\theta$ in the direction  $\nabla_\theta \log \pi(a_{t_k}|s_t; \theta)$. The update step size is $\alpha(r_t - b(s_t))$.
Otherwise, the solution probability is decreased. In this way, we reinforce actions that lead to better
rewards. We use entropy factor $\beta=0.1$ (strength of the entropy regularization term) in the training phase in order to ensure that the DRL agent explores the action space adequately to discover good policies.

\myparab{Implementation:}We develop our deep reinforcement learning system in Python 3. The current prototype uses TensorFlow \cite{abadi2016tensorflow}. Our policy neural network has three layers. The first layer is a convolutional layer with 128 filters. The kernel size is 3 × 3 and the stride is set to 1. The second layer is a fully-connected layer with 128 neurons. We use Leaky ReLU for the activation function for the first two layers. The final layer is a fully connected linear layer (without activation function) with $N*(N - 1)$ neurons corresponding to all possible sets of flows in the network. The softmax function is applied upon the output of the final layer to generate the probabilities for all available actions. The learning rate $\alpha$ is initially configured to 0.001 and decays every 500 iterations with a base of 0.96 until it reaches the minimum value of 0.0001. We use python LP modeler, pulp, \cite{pulp} to find the optimal solution of (\ref{cons:optimal2}).

\section{Evaluation}
\label{experiment}
\vspace{-0.03in}
\eat{We multiply the demans of the Abilene network with 10 and of the ATT by 60 and divided the link capacity of ATT by 10.}

In this section, we experimentally evaluate our scheme and compare it to other classical TE approaches.  Our goal is to understand the performance of the DRL approach in identifying robust paths at different times in order to minimize network update costs while guaranteeing low link utilization.

\myparab{Network topology:} In our evaluation, we use Abilene and ATT network routing topologies. The number of nodes in the Abilene network is 12 and the directed links of the network are 30. For this network, the measured traffic matrices and network topology information (such as link connectivity, weights, and capacities) are available in \cite{Abilene}. ATT network has 25 nodes and 112 links. We obtained the topology and data set for the ATT topology from \cite{bogle2019teavar}. Since the link capacities for ATT topology are not available, we randomly set the link capacity in the range of 1,000,000 and 9,000,000 bps for each link. The weight of each link is set to one. 

\myparab{The workload}: The data set of Abilene and ATT network is for 7 and 2 days respectively. The demand matrices are measured with the granularity of 5 and 15 minutes for Abilene and ATT networks respectively. In order to stress the workloads to avoid small link utilization we multiply the traffic of each flow in each traffic matrix by 4 and 8 in the workload, for network Abilene and ATT respectively. In our experiments, we randomly choose 70\% of our dataset as the training set for our scheme and the remaining 30\% of our dataset as a test set for testing all schemes.

\myparab{Algorithms:}We compare our scheme with oblivious routing, ECMP, MLU-optimal, and the optimal scheme (the solution fo the optimization problem (\ref{cons:optimal2})). We assume the optimal scheme knows future demands and can identify the paths that should be used in future time intervals beforehand. While the optimal scheme knows the future demands, it is an approximation of the optimal solution because of the non-convexity of the problem for finding optimal paths (explained in section $\S$ \ref{sec:optimal_solution}). The MLU-optimal Scheme does not care about the update cost and only minimizes the maximum link utilization in each time interval by solving the multi-commodity flow problem. In each topology, we set the value of $|R|$ to the average number of paths used by the MLU-optimal scheme over all the 5-minute time intervals.

\begin{figure}

  \begin{subfigure}{.22\textwidth}
    \includegraphics[width=4.0cm]{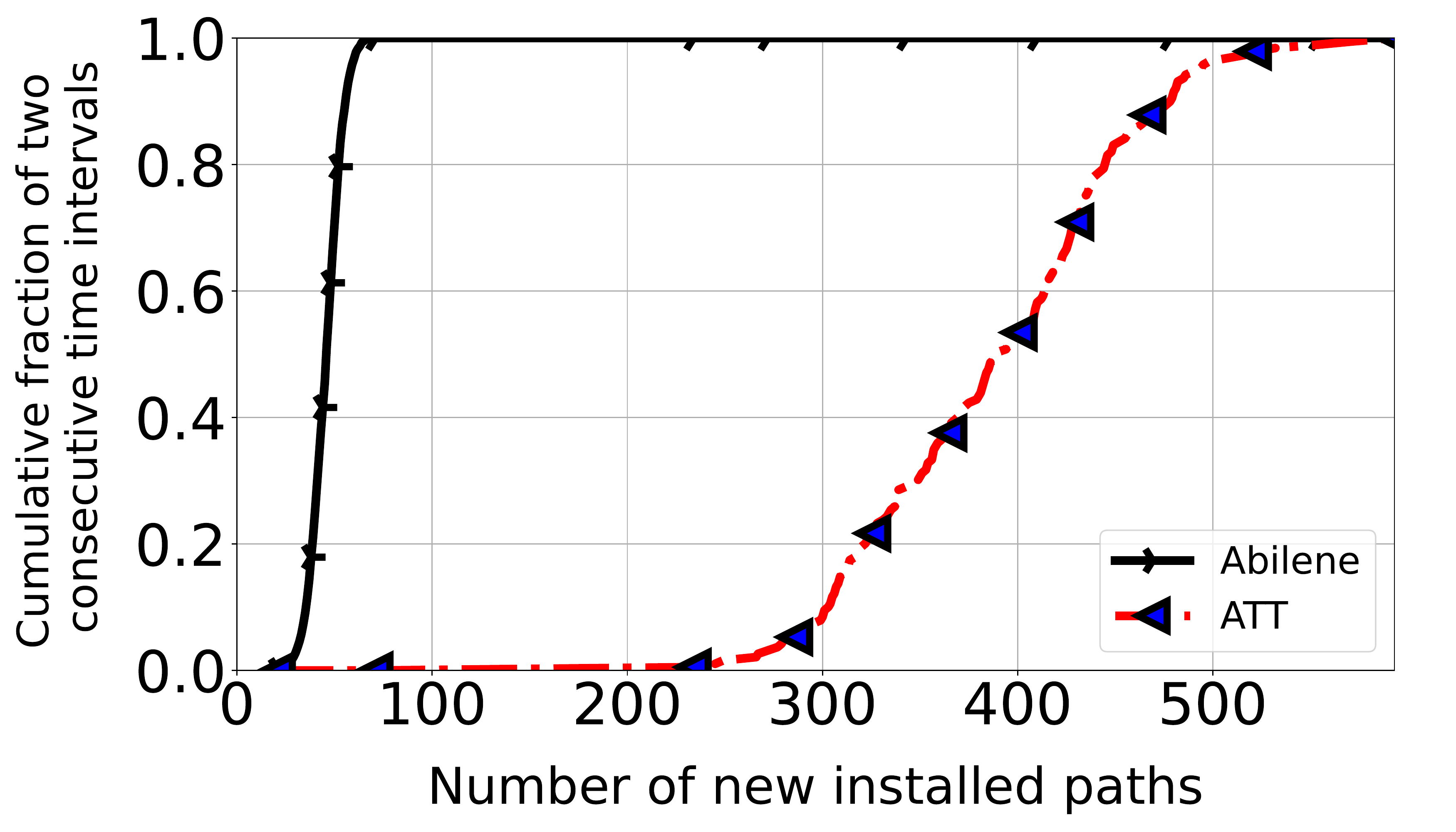}
    \vspace{-0.2in}
  \caption{Path updating frequency \label{fig:path_updating_cost}}
  \end{subfigure}
  \begin{subfigure}{.22\textwidth}
    \includegraphics[width=4.0cm]{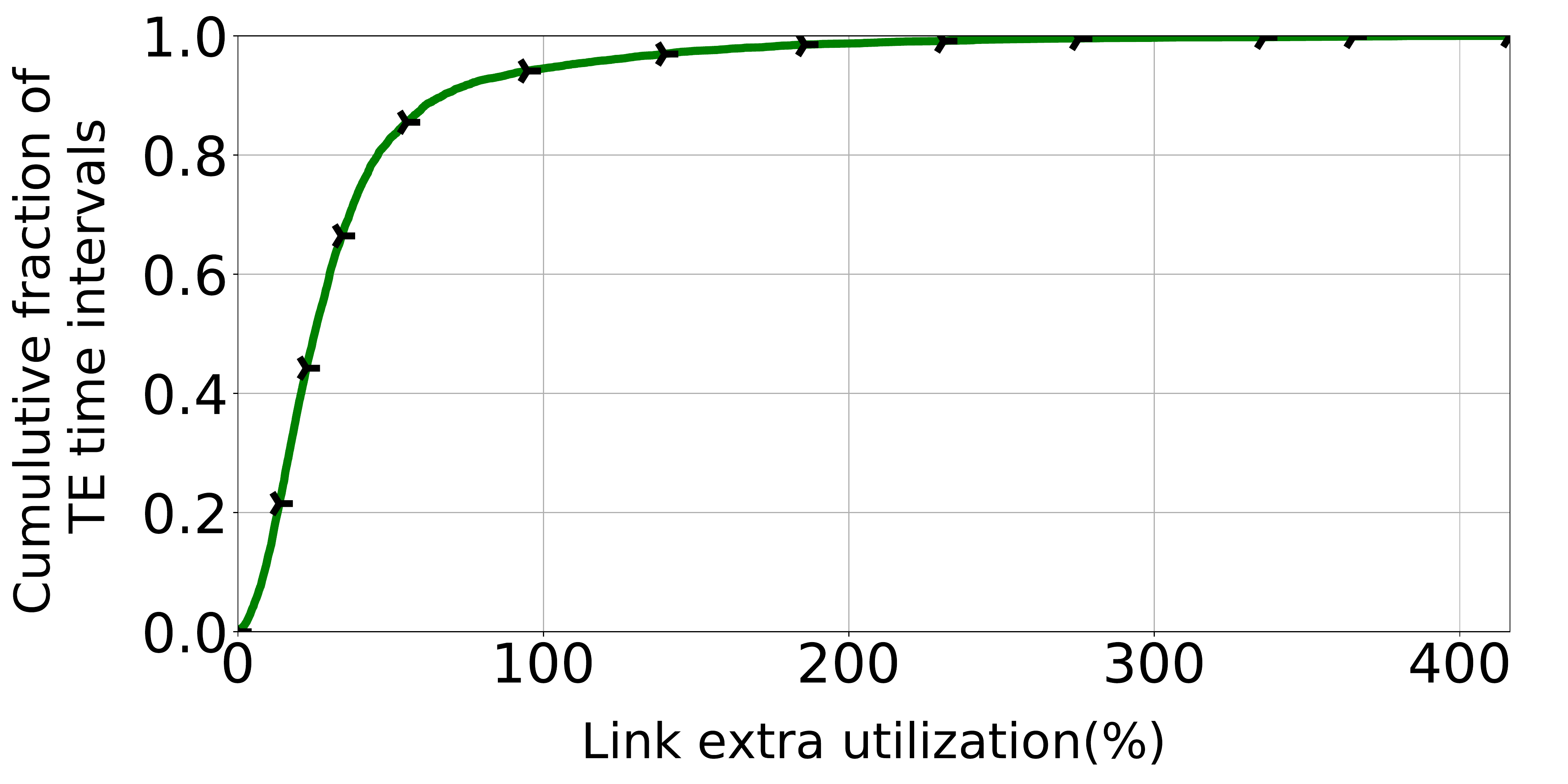}
    \vspace{-0.2in}
    \caption{Link over-utilization during route computation \label{fig:link_over_utilization2}}
  \end{subfigure}
\vspace{-0.06in}
  \caption{MLU-optimal scheme suffers from a high path updating overhead (\ref{fig:path_updating_cost}). In addition, a slow reaction to changes in traffic matrices results in high link utilization during route computation in this scheme (\ref{fig:link_over_utilization2}).  \label{fig:mlu_optimal_overhead}}
\end{figure}

\myparab{Frequency of path updating:} We measure the difference between the set of paths in the solution of the multi-commodity flow problem of two consecutive time intervals. We measure the average number of paths that change across all flows in the network. Figure \ref{fig:mlu_optimal_overhead}.a shows that in 50 percent of two consecutive time intervals, we have at least 300 path updates in the ATT network and 5 or 6 paths for the Abilene network. The reason for this difference is that the ATT network has more edges and more flows. This shows the high overhead of updating the forwarding behavior of routers in networks with a huge number of nodes and edges. In the next section, we show that our scheme is able to avoid the overhead of path updates at the cost of slightly larger MLU in comparison to the MLU-optimal scheme.

Minimizing the MLU by solving the multi-commodity flow problem is not scalable and can take minutes for large networks \cite{abuzaid2021contracting,zhu2021network}. During  route computation using this approach, routers would use the currently installed paths to route the new traffic. Figure \ref{fig:mlu_optimal_overhead}.b shows the link utilization of the network using the old paths while the centralized controller solves the multi-commodity flow problem for the MLU-optimal scheme. The \textit{y-axis} shows the difference in MLU of links using the old and new paths. We can see that most of the links can experience a difference of 80\% higher link utilization using the old paths in comparison to using new paths.

\eat{We can see that the hindsight approach that knows the exact DM in the next $w$ intervals also gives us 25 percentage MLU benefit. In another word, our agent is able to find the set of the paths that are almost as the optimal one.}

\begin{figure}

  \begin{subfigure}{.22\textwidth}
    \includegraphics[width=4.0cm]{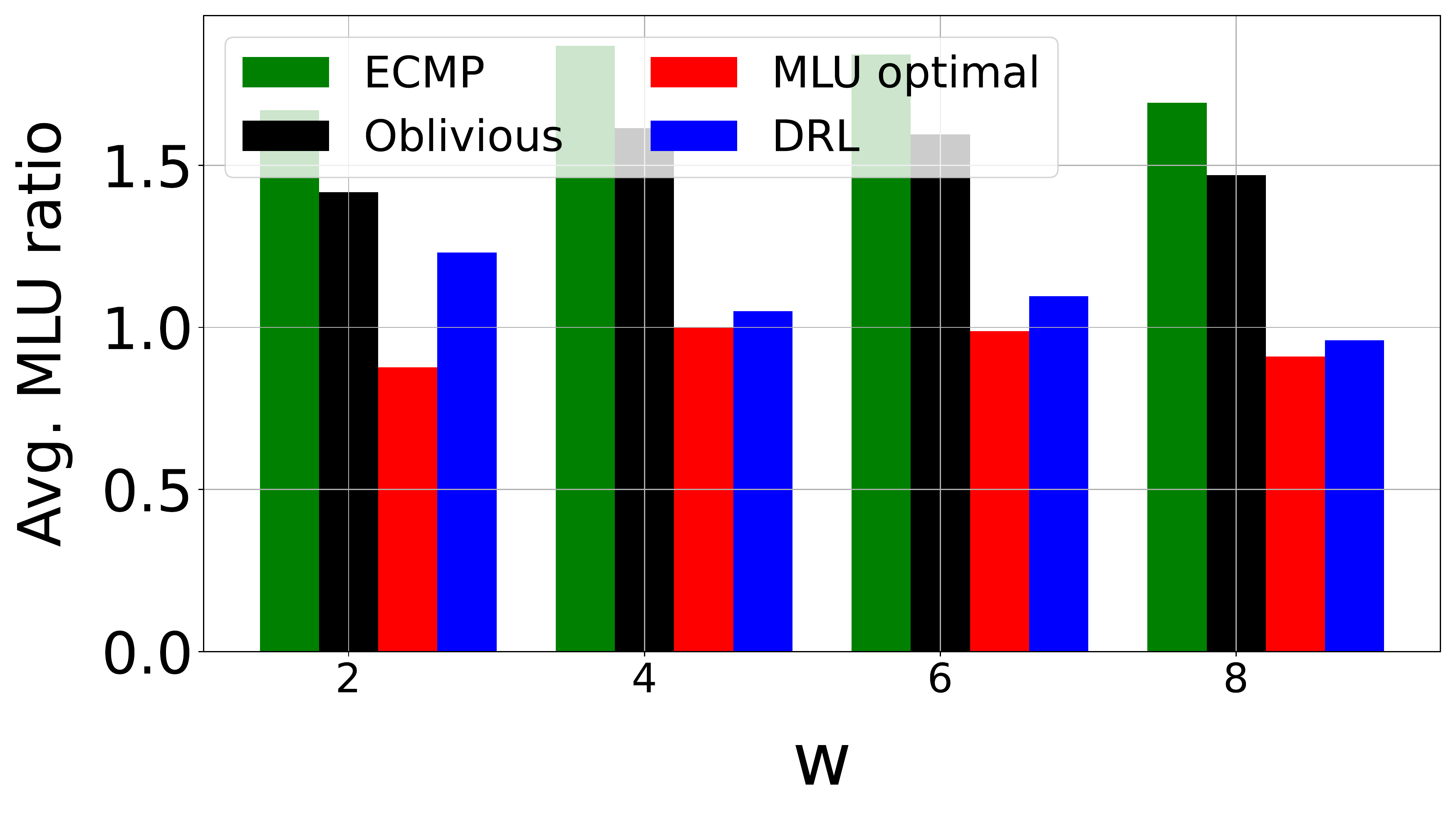}
    \vspace{-0.2in}
  \caption{ATT \label{fig:att}}
  \end{subfigure}
  \begin{subfigure}{.22\textwidth}
    \includegraphics[width=4.0cm]{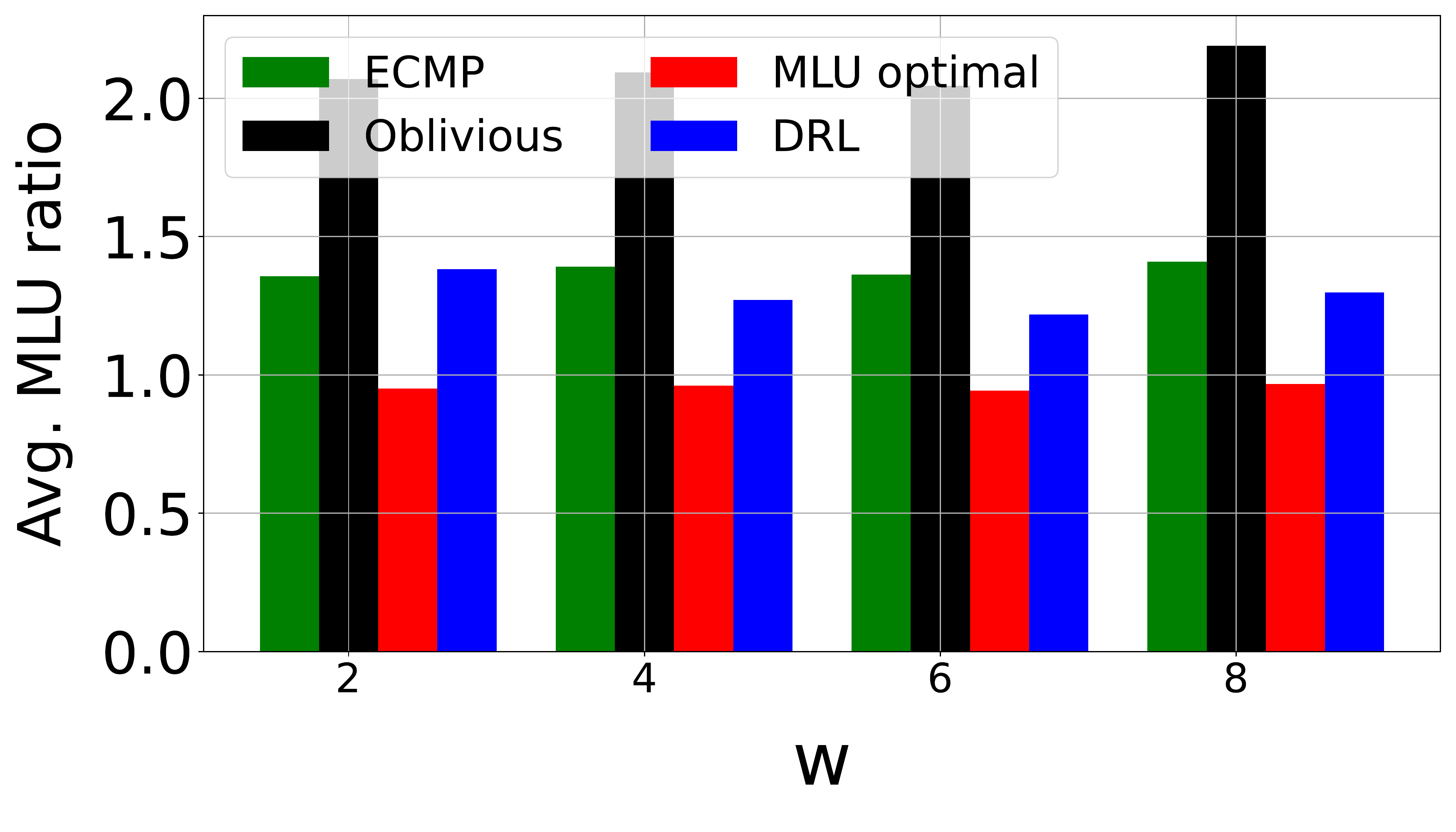}
    \vspace{-0.2in}
    \caption{Abilene \label{fig:abilene}}
  \end{subfigure}
\vspace{-0.06in}
  \caption{Our DRL-based scheme is able to provide a lower average MLU ratio (to the MLU in the optimal scheme) compared to other basic TE schemes. The MLU-greedy scheme achieves a lower MLU ratio but with a cost of updating paths in each time interval (fig. \ref{fig:path_updating_cost}). The reason to have a ratio value less than 1 is because of our relaxation of constraints to find the optimal paths ( Section $\S$ \ref{sec:optimal_solution}). \label{fig:affect_of_window_on_MLU}}
\end{figure}

 \begin{figure}[t]
\centering
    \includegraphics[width=5cm]{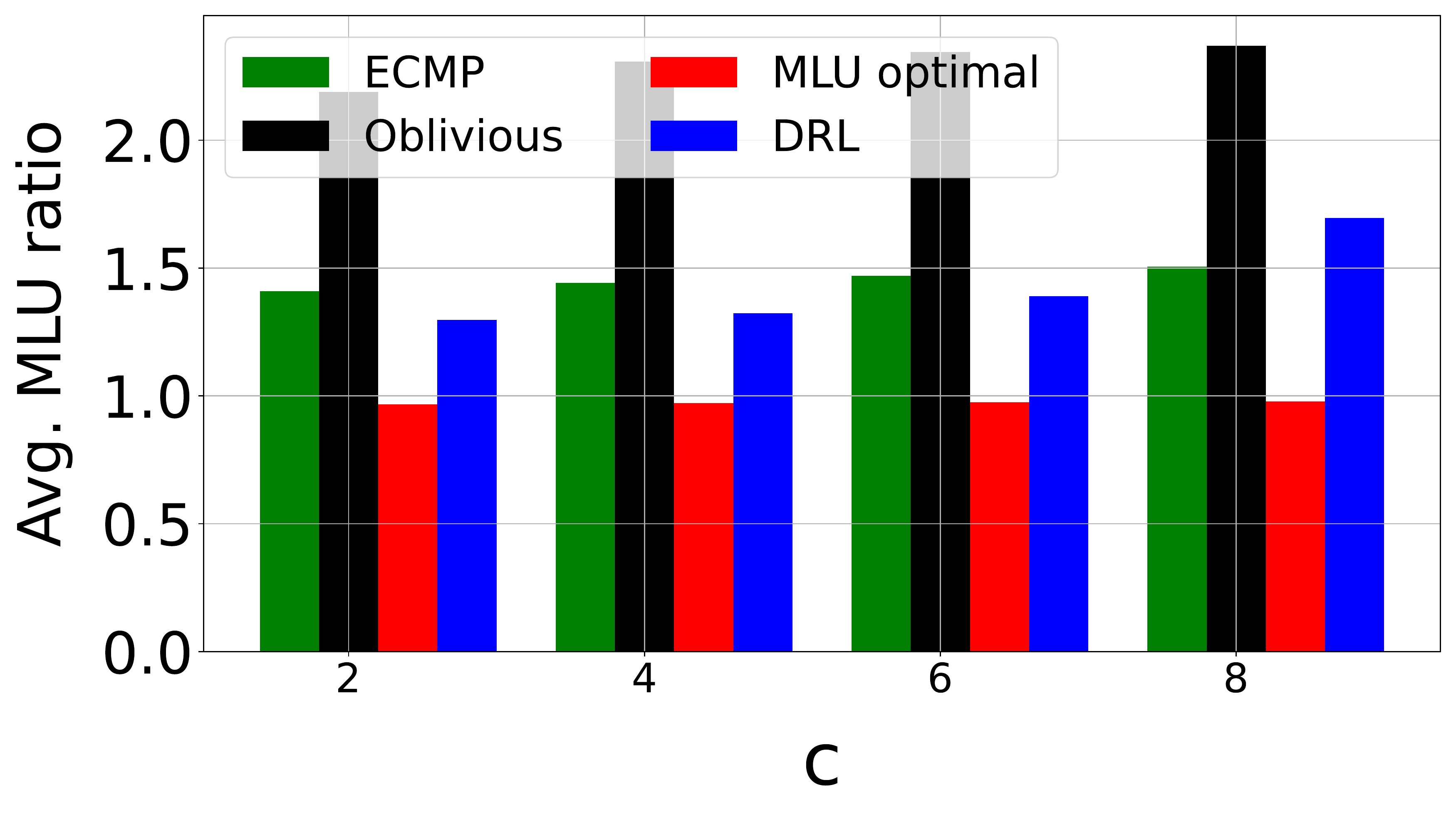}
    \vspace{-0.15in}
  \caption{Average normalized MLU by the MLU of the optimal-MLU wise as a function of $c$ .\label{fig:affect_of_c_window_on_MLU}}
\end{figure}

\myparab{Average link utilization:} In this experiment, we compare link utilization in our scheme to link utilization using classical TE schemes such as ECMP and Oblivious routing. We train the DRL agent separately for different values of $w$ for look ahead window size.  The value of the commitment window size is $2$. Figure \ref{fig:affect_of_window_on_MLU} shows the average ratio of MLU (to the MLU in the optimal scheme) using different schemes. The result for our DRL-based scheme is for 1,000 epochs of training. Figure \ref{fig:affect_of_window_on_MLU} shows that our scheme can outperform the ECMP and the oblivious routing scheme. The MLU-optimal scheme that only cares about link utilization provides a lower MLU ratio. The \textit{MLU-optimal} scheme only cares about the link utilization and provides a lower MLU ratio but of the cost of traffic fluctuation and path updates on the routers in each time interval. It also suffers from high route computation time. Our scheme is efficient in terms of computing a new solution as it can be trained offline. Testing the neural network with a new state to compute a new solution (a set of paths) can be done in less than one second.

Figure \ref{fig:affect_of_c_window_on_MLU} shows the average ratio of MLU (MLU in our scheme to the MLU of the optimal scheme) during one day for different values of $c$ for the Abilene network. Results are for 1,000 epochs. Intuitively, the larger the observed time window $c$, the more information the DRL agent can rely on, and the higher the accuracy of the traffic behavior changing within the next time windows. However, we observe that a higher value of $c$ may not necessarily result in better path selection. We show in the next section that different values of $c$ requires different number of epochs for training to converge. It is possible that higher number of epochs of training for the higher $c$ may result in lower MLU ratio. We consider this as a future work.

\begin{figure}

  \begin{subfigure}{.22\textwidth}
    \includegraphics[width=4.0cm]{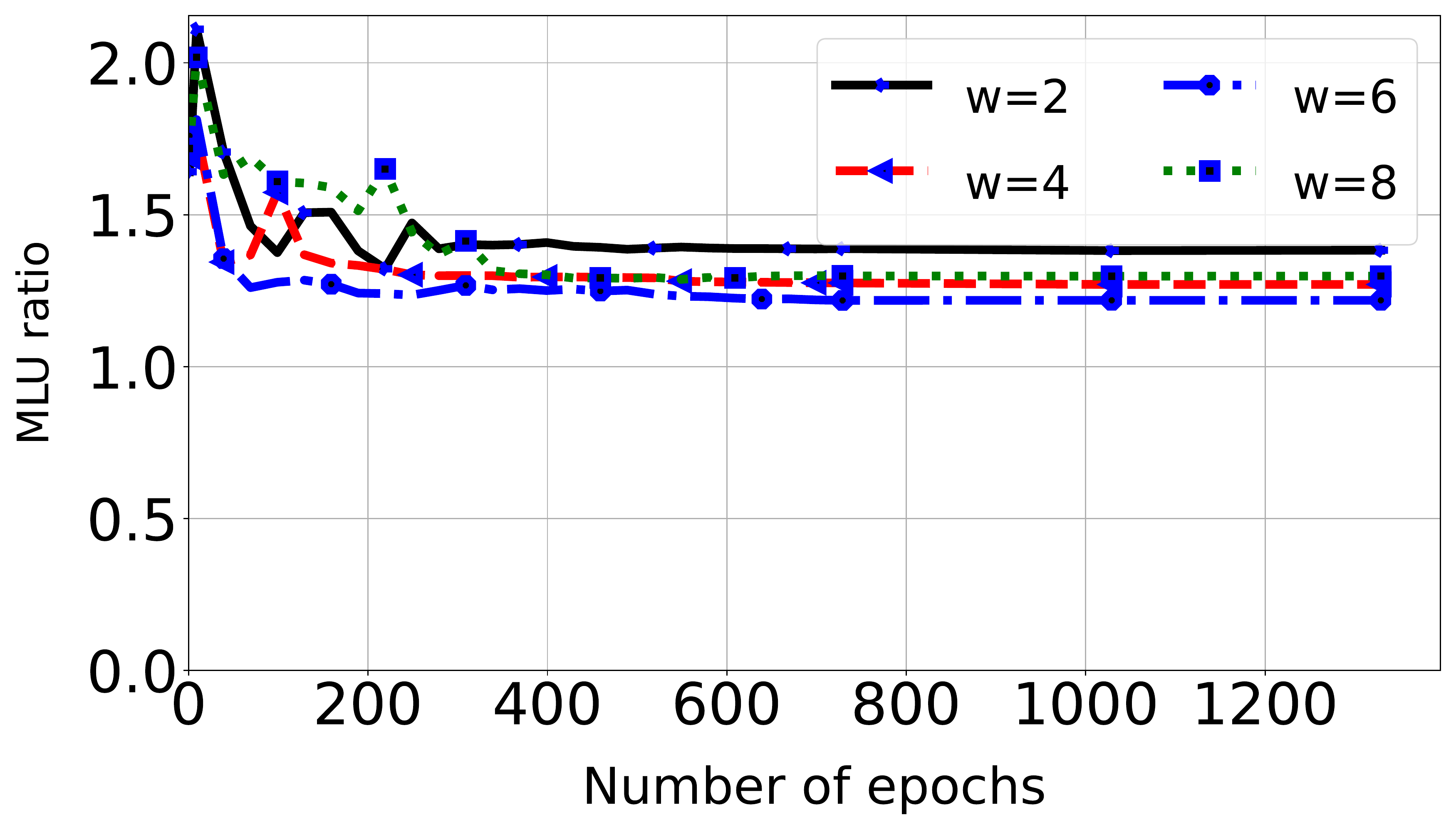}
    \vspace{-0.2in}
  \caption{Impact of $w$ on convergence. \label{fig:sensitivity_analysis_as_w}}
  \end{subfigure}
  \begin{subfigure}{.22\textwidth}
    \includegraphics[width=4.0cm]{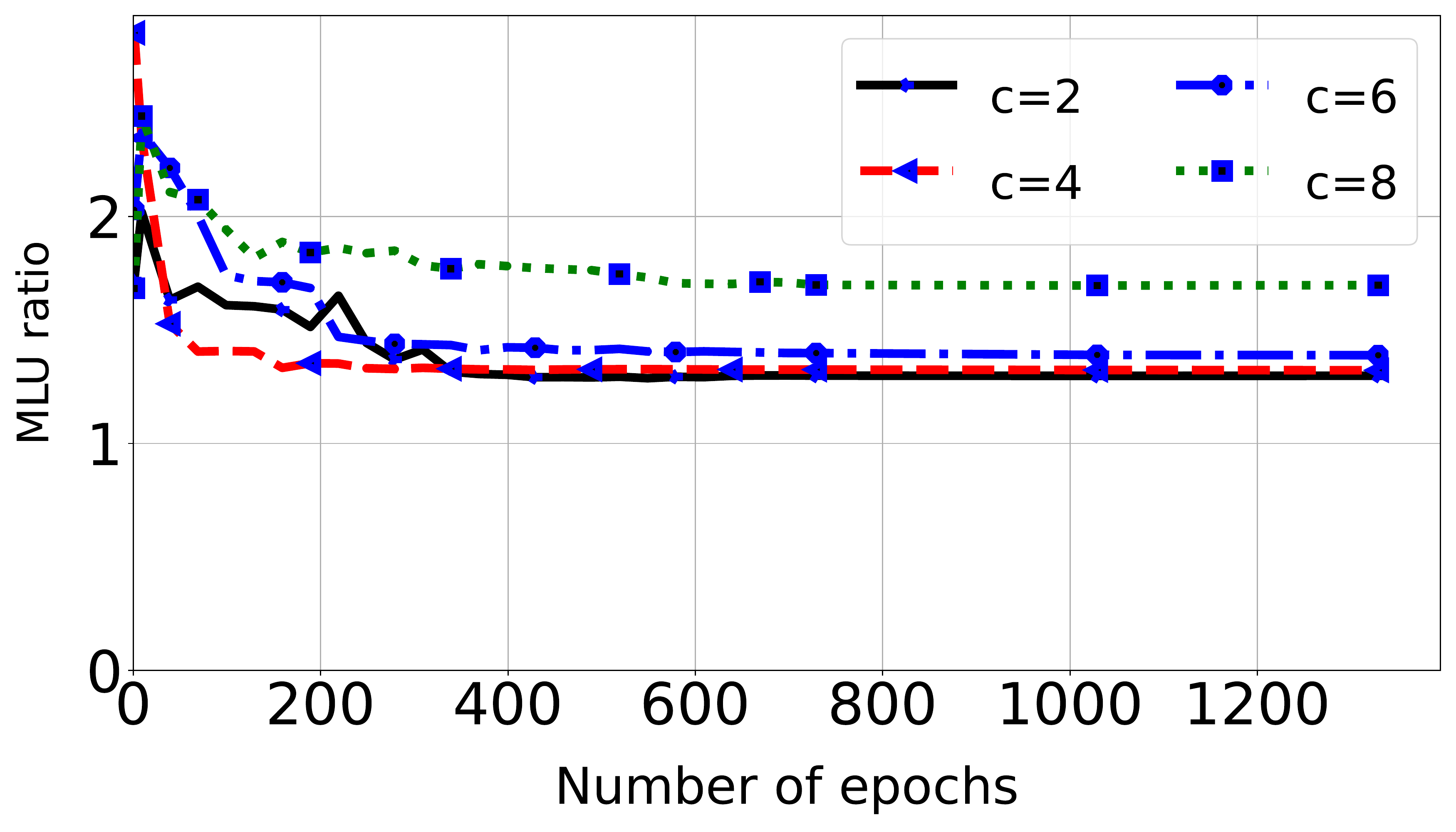}
    \vspace{-0.2in}
    \caption{Impact of $c$ on convergence. \label{fig:sensitivity_analysis_as_c}}
  \end{subfigure}
\vspace{-0.06in}
  \caption{Impact of $w$ and $c$ on convergence speed. The MLU results are normalized with the optimal scheme. \label{fig:sensitivity_analysis}}
\end{figure}

\myparab{Sensitivity Analysis:} To better understand the impact of  parameters $w$ and $c$ on the convergence of the training algorithm of our scheme, we conduct a sensitivity analysis. In Figure \ref{fig:sensitivity_analysis}, we show the MLU ratio (to the MLU in the optimal scheme) using our scheme to the optimal solution as a function of  the number of epochs for different values of $w$ and $c$. Each epoch in our training algorithm corresponds to a simulated one-day period. We see that with lower values of $w$, the DRL agent solution converges quickly, while with higher values of $w$ we have high variance and the algorithm converges quite slowly. This is because a more complicated relationship needs to be considered  between previous time intervals and the paths that should be used in future time intervals and this can be trained in more epochs.

\label{evaluation}

\section{Related Work}
\vspace{-0.02in}
Recently researchers have tried to use learning-based solutions for different networking problems \cite{chen2018auto,mao2019learning,mao2017neural,chen2022rl}. In this section, we overview the research that has been done in the field of using machine learning for TE. 


\myparab{Learning for TE:} The idea of using learning techniques for TE is not a new topic and this idea has been proposed in the late 90s in some work such as \cite{peshkin2002reinforcement,boyan1994packet}. Asef et. al \cite{valadarsky2017learning} use DRL to learn a map from previous $k$ DMs to the next DM and routing configuration. Zhiyuan et. al \cite{xu2018experience} uses DRL to learn the split rates of the traffic on a set of established paths in the network in order to minimize the end-to-end delay in the network. A multi-agent RL approach has been proposed in \cite{geng2020multi} for distributed TE. It uses multiple agents in each region of the network to control the terminal
traffic and outgoing traffic for each region. CFR-RL \cite{zhang2020cfr} uses DRL to select a small set of flows and then only reroute these flows in the network in each time interval using a fixed set of paths. A similar idea has been proposed in \cite{sun2020scalable} that uses DRL to select a subset of links (critical links) in the network and dynamically adjust the link weights for
the critical links in order to minimize the end-to-end transmission delay.

An online learning-based approach has been proposed in \cite{zheng2021online} where the cost of network updating and other objectives such as link utilization are considered jointly. However, online-learning algorithms can only guarantee the upper bound for competitive ratio in the worst case (in comparison to the optimal scheme) while we can have better data-driven algorithms such as RL for the problem. The proposed idea of using multiple agents for TE  in \cite{geng2020multi} can be used in our setup to select a different set of paths at different regions of the network. 


\myparab{Oblivious routing:} Oblivious routing approaches do TE without considering the DM in their route computation. Using Räcke's oblivious path selection algorithm, SMORE \cite{kumar2018semi} uses a fixed set of paths to avoid the path-changing problem in the network. However, a fixed set of paths may not provide the optimum solution for each time interval. Our scheme also tries to find a robust subset of paths that should be used in the network in multiple future time intervals. 


\section{Conclusions}
\label{sec:concl}
\vspace{-0.06in}
Inspired by recent successes of DRL techniques in solving complex online control problems, in this paper, we proposed a data-driven algorithm that does the path selection in the network. We deployed and evaluated our approach on a real testbed, and demonstrated its performance and adaptiveness to traffic demands for two real network topologies and datasets. A huge benefit of our approach comes from its ability to limit establishing new paths in the network over time and make the decision for path selection quickly without having to solve an LP problem in each time interval. In different experiments, we are able to show that our approach performs better than classical approaches such as ECMP and oblivious routing and can achieve the near-optimal solution.

\bibliographystyle{plain}

\bibliography{sample-bibliography} 
\vspace{-0.2in}

\end{document}